\documentclass[conference]{IEEEtran}
\IEEEoverridecommandlockouts
\usepackage{cite}
\usepackage{amsmath,amssymb,amsfonts}
\usepackage[noend]{algpseudocode}
\usepackage{algorithmicx,algorithm}
\usepackage{graphicx}
\usepackage{textcomp}
\usepackage{xcolor}
\usepackage{mathtools}
\usepackage{pifont}
\usepackage{booktabs}
\usepackage[noend]{algpseudocode}
\usepackage{mathtools}
\usepackage{pifont}
\begin{document}
\title{PRETRUST: A Framework for Fast Payments in Blockchain Systems\\
\thanks{This work was funded by R\&D Program of Beijing Municipal Education Commission with grant number:110052971921/021.
	
This work has been submitted to the IEEE for possible publication. Copyright may be transferred without notice, after which this version may no longer be accessible.}
}

\author{\IEEEauthorblockN{1\textsuperscript{st} Huapeng Li}
\IEEEauthorblockA{\textit{School of Information} \\
\textit{North China University of Technology}\\
Beijing, China \\
bwvanh114@outlook.com}
\and
\IEEEauthorblockN{2\textsuperscript{nd} Baocheng Wang}
\IEEEauthorblockA{\textit{School of Information} \\
\textit{North China University of Technology}\\
Beijing, China \\
wbaocheng@ncut.edu.cn}
}

\maketitle

\begin{abstract}
In this study, we propose PRETRUST, a new framework to address the problem of the efficiency of payment process based on blockchain systems. PRETRUST is based on the thoughts of consortium chains, supporting fast payments. To make parties address transactions efficiently and scale out the capacity, the main strategy is based on the guarantee mechanism and shard technology with a random arrangement of transactions and consensus groups. In PRETRUST, the guarantee process replaces the verification of a transaction on-chain to realize the optimization of efficiency. And the model of a trusted execution environment is utilized in the interactions between PRETRUST and public systems. Throughout the paper, we discuss several transaction scenarios and analyze the security, which shows that PRETRUST can guarantee security.
\end{abstract}


\section{Introduction}
In 2008, Satoshi Nakamoto published ``Bitcoin: A Peer-to-Peer Electronic Cash System", which describes the framework of the cryptocurrency system based on blockchain technology  \cite{nakamoto2008peer}. Originating from Bitcoin, blockchain technology is a comprehensive product of distributed data storage, point-to-point transmission, consensus mechanism, encryption algorithm, and other computer technologies. Blockchain has been paid more and more attention as one of the underlying technologies of Bitcoin. The main advantage of blockchains lies in the decentralized design. Through the leverage of the encryption algorithm, consensus mechanism, and reward mechanism, the point-to-point transaction can be realized in the distributed network where nodes do not need trust. In 2013, Vitalik Buterin released the first edition white paper of Ethereum. Compared with Bitcoin, the most important function of Ethereum is to add the smart contract and Ethereum reduces the time of generating a block greatly from ten minutes to ten seconds. In the era of blockchains, Ethereum, Corda, and Zcash \cite{miers2013zerocoin} and many other novel frameworks (e.g., \cite{kokoris2018omniledger,churyumov2016byteball,silvano2020iota}, and so on) and flexible applications (e.g., \cite{nathan2020blockchain,devine2021conceptualising}, and so on) are emerging.

A blockchain network provides a secure and reliable way to record data and execute smart contracts, but it has problems with efficiency. Taking Bitcoin as an example, every transaction will be published to each node, and nodes will verify it with the records and pack them into a new block. Then the miner will obtain the opportunity to publish the block by providing computing power proof and get the block reward. After a certain number of consecutive blocks by nodes of the whole network, the block with the computing power proof becomes a part of the blockchain system. There are mainly two parts involved in this process, one is the miner computing the hash value of the block and competing for the privilege of publishing the block, and the other is waiting for the effective confirmation of the block on the chain. The cryptography mechanism ensures the security and reliability of block records, but it is at the cost of efficiency \cite{garay2015bitcoin,kiffer2018better}, This problem has become one of the main reasons hindering the application of blockchains, especially in transactions calling for fast payments.

It is necessary to find a solution to improve the efficiency problem of blockchains for fast payments. Here, we consider a common application scenario as an example. In the daily small transactions, the buyer and the seller have the immediate requirement for the confirmation of the transaction. However, many payment methods based on blockchains cannot support rapid confirmations of payments. In detail, the structure of the blockchain itself limits the scalability of the structure of blockchains. In the traditional single chain structure blockchain system, to get the final consensus and the security of the system, the speed of packing a block cannot be accelerated indefinitely. Moreover, to pursue greater benefits, miners prefer to package transactions with Upper transaction costs into their blocks. In this case, the interaction of daily transactions and the blockchain system is one of the main reasons that cause the native blockchain system to fail to ensure efficiency. To improve the throughput of blockchain, from the perspective of changing the structure of blockchains, the mainstream technology is sharding \cite{kokoris2018omniledger}, \cite{luu2016secure} and DAG (Direct Acyclic Graphs) \cite{churyumov2016byteball}, \cite{silvano2020iota}; From the perspective of transferring workload, off-chain capacity scale-out technology can be adopted without changing the underlying structure. In contrast and the latter is not the design of blockchains in essence, but focuses on a protocol executed off-chain \cite{poon2016bitcoin,khan2019lightning,kaur2020scalability,yan2020confidentiality}. The redesign of blockchains is a complex project facing many challenges. To be compatible with the existing mature blockchain frameworks for instant payments, considering the protocol design outside the public chain is a flexible and effective strategy.

The paper proposes a new framework: PRETRUST, which can improve the transaction efficiency to respond to the demand for instant payments. We suppose a transaction scenario of $Alice$ and $Bob$: $Alice$ needs to pay for $Bob$ to get a commodity. The general idea is that $Alice$ first need a trusted miner to get a payment guarantee, and then shows it to $Bob$. $Bob$ can provide a commodity to $Alice$ after confirmation of the guarantee. $Alice$ needs to ask a “guarantor” to provide a guarantee for each transaction, that is,  $Alice$ getting the payment guarantee. With this method, the transaction between the buyer $Alice$ and the seller $Bob$ will be accomplished with the participation of a guarantor. Formally, there exists no need for $Alice$ and $Bob$ to participate in the waiting process of the transaction being confirmed in the chain anymore while the rationality and security are realized by PRETRUST, which is designed off-chain. In this way, this part of the time can be saved to improve the transaction efficiency. The design and analysis of PRETRUST are described in detail in Section \ref{des} and Section \ref{protocol}.

PRETRUST is built on the existing public blockchain systems, forming a structure of side chains. To distinguish, this paper makes the former an external environment and the latter an internal environment. For clear description, part of the side chain is denoted by the “$internal\ environment$” while the public blockchain systems which PRETRUST relies on are denoted by “external environment”. The internal environment adopts shard technology with a random division of consensus committees and transactions, which will be described in Section \ref{des}. All parties involved in PRETRUST need to pay tokens to the contract account in the external environment and then obtain the equivalence they have in the internal environment through the records of the public chain. In PRETRUST, every miner in the external environment blockchain network can become a guarantor. As the guarantor, on the one hand, he needs to process guarantee requirements from other clients; On the other hand, he needs to record and supervise the information of other consensus committees. To achieve these two main purposes, the guarantor needs to pay a certain amount of deposit in the external environment in advance for supervision right and guarantee qualification. And the main objectives and contributions of this paper are as follows:

\textbf{Efficient provision of guarantee certification:} It is the responsibility of the payer to produce a valid payment certificate in a transaction. However, the existing blockchain trading system hinders the efficiency of blockchain networks, so the payee must wait long enough to confirm the transaction. Therefore, the main motivation of this paper is to save the long confirmation waiting time in traditional blockchain systems by constructing a series of rules between the payer and the payee. To achieve this goal, based on the existing blockchain frameworks supporting smart contracts, we try to select a trusted guarantor outside the chain to avoid waiting. Under this assumption, the guarantor needs to verify the effectiveness of the transaction of the payer and lock the amount of this transaction before providing an effective guarantee certificate. Therefore, in the process of transaction, the payee can provide the committees to the payer on the assumption of trusting the guarantor. To effectively realize the security of this guarantee mechanism, this paper designs the transaction processing and supervision mechanism respectively.

\textbf{Dynamic consensus groups division:} We adopt the idea of shard technology to get high efficiency and security at the same time. The sharding technology in a blockchain network is a structural scheme for performance, which can be roughly divided according to the attributes of the network region and the transaction itself. Different transactions are processed by category to improve the throughput of the blockchain. In PRETRUST, the construction of the internal environment adopts the idea of shard technology to improve the processing speed of transactions. The specific consensus group has the right of bookkeeping the shard. Different transactions will be assigned to different consensus groups for recording according to different payers. To improve security, members of a consensus group in the internal environment need to be randomly assigned regularly.

\textbf{Punishment mechanism of mutual supervision:} We designed a supervision mechanism among consensus groups to further ensure the security of the protocol. In the internal environment, each consensus group is responsible for the blocking of the shard. Each transaction is assigned to the corresponding consensus group according to the attribute of the payer, which is responsible for the bookkeeping and guarantee for the payer. Each guarantor is also under the supervision of the corresponding consensus group. When a guarantor is found adversary, the corresponding consensus group in charge of supervision needs to issue effective certificates to punish him with proof from $Alice$.

\section{Related works}
PRETRUST is designed mainly with the idea of a guarantee mechanism to improve addressing efficiency, which is an off-chain expansion scheme designed with shared technology. As mentioned above, public blockchains supporting smart contracts are called the external environment, and the interaction conditions that rely on the public chain are called the internal environment. Consequently, the design of PRETRUST is divided into two parts: Internal and external interactions and corresponding transaction rules. And these two parts are related to the theory and technology mentioned below:

\textbf{Scale-out:} The main idea of off-chain expansion \cite{gudgeon2020sok,worley2018blockchain} is to transfer part of the data to the off-chain for processing and send the results to the public chain for other addressing \cite{hepp2018chain,muhlberger2020foundational,chang2019supply,reijers2021now}. According to different transfer modes, there are mainly state channels \cite{coleman2018counterfactual,miller2019sprites,mizrahi2021congestion}, and side chains \cite{back2014enabling,singh2020sidechain}. Taking Lightning Network as an example, the dominant parts of the off-chain structure are not the elements of blockchains, but the next transmission network without any relation to public blockchains. In the Lightning Network, both users need to deposit tokens in blockchain systems which are relied on to establish a transaction channel in advance and jointly maintain a ledger.  After building up a channel of the two parties, transactions are all handled in the off-chain channel and only submit the final status to the chain for updating the status records before closing the channel when they intend to withdraw their tokens deposited at the beginning. The main contribution of this kind of framework is that except for the final transaction, other transaction processes are not recorded in the blockchain when the channel needs to be closed, And the efficiency of the system is improved with this method. To facilitate the transaction between the two parties who have not established a channel, Raiden Network \cite{hees2016raiden} supplements and expands Lightning Network. Raiden Network realize the transactions between the two parties who have not established a channel through the interaction with a third party, an intermediary indeed, which has already established channels shared with the two parties respectively. Lightning Network and Raiden Network effectively reduce the interaction with the public chain to a certain extent, but due to the establishment and closure of the channel, malicious participants can lock the deposit of parties who have built up channels for a long epoch. With the method of not cooperating with the others after establishing the channel, malicious participants attack the system at a relatively low cost. Besides, the addition of intermediaries will also lead to the centralization of the system to a certain extent and affect security.

\textbf{Shard technology:} $Shard$ means that nodes are divided into several smaller consensus groups to process transactions in parallel and maintain independent blockchain records. ELASTICO \cite{luu2016secure} is the first shard protocol designed for public chains with the strategy of shard transactions. The main idea of ELASTICO is to divide the nodes into several small committees, and each committee will deal with disjoint transaction sets. These disjoint transaction sets are called shards. The node establishes identity information by proof of work, and then the node is randomly assigned to each committee. The Byzantine consensus protocol \cite{castro1999practical} runs within the committee, and they process transactions in parallel. However, PBFT protocol has too many communication rounds with high communication complexity, which is one reason for the delay in the protocol. To improve the efficiency, OmniLedger \cite{kokoris2018omniledger} builds on the Byzantine consensus scheme in ByzCoin \cite{kogias2016enhancing}, because it scales efficiently to thousands of consensus group members. Omniledger is composed of an identity chain and multiple shards. Nodes are automatically assigned to different shards through the Rand Hound protocol. To avoid the centralization risk caused by a specific consensus group that holds the recording right of recording the corresponding shard for a long term, Omniledger adopts the strategy of epoch organization but does not propose an efficient interaction strategy among shards. And that brings a lot of consumption of data migration. SsChain \cite{chen2019sschain} divides the blockchain network into a root chain network and a shared network. And the incentive mechanism dynamically adjusts the computing power to make the computing power evenly distributed in different shards. Sschain adopts the market incentive mechanism, so it does not need to reorganize the shard network regularly and avoid data migration consequently with this design. Distinctive from the dynamical shard mode just mentioned, a static shard is proposed \cite{yoo2018blockchain}. The model divides the blockchain network into domain shard and global shard. There is a committee determined by proof of work in the domain partition which is adjusted dynamically. Nodes in the committee verify transactions and generate blocks through PBFT consensus algorithm. The committee of global shards is also determined by proof of work, and the committee verifies the transaction through PBFT. After the transaction passes the verification, it is transmitted to the corresponding domain shard. The nodes in the domain shard are responsible for generating blocks and recording them to the shard. This scheme provides the idea of shard by levels, but the static shard strategy will lead to the risk of centralization.

\textbf{Leverage of TEE:} Trusted execution environment is a concept of secure hardware \cite{sabt2015trusted}. This environment can ensure that the computation will not be disturbed by the malicious operation, so it is called "trusted". Many works apply TEE in blockchain technology \cite{lind2018teechain,zhang2016town,milutinovic2016proof,chen2017security}. Teechain \cite{lind2018teechain} establishes an off-chain payment channel with the leverage of TEE. Town Crier \cite{zhang2016town} acts as a bridge between smart contracts and existing websites, which are already commonly trusted for non-blockchain applications. Proof of Luck \cite{milutinovic2016proof} and Proof of Elapsed Time \cite{chen2017security} use TEE to provide randomness to ensure that the consistency of consensus is not affected by network conditions. In this PRETRUST, to guarantee the security of interactions between the internal environment and the public chain system when any party intends to withdraw the tokens in the internal environment, TEE is responsible for outputting effective certificates as the criterion of withdrawing.

To better support the current mainstream blockchain projects, PRETRUST is designed to be built on the public chain, which is a kind of off-chain scaling-out strategy. Based on the technology and theory above, PRETRUST adopts the idea of the guarantee certification to take place the on-chain certification in normal blockchain systems. Unlike the reconstruction of the public chain, the application of PRETRUST protocol will not change the system of the public chain. Therefore, it can be built on various public chain projects supporting smart contracts, with the general advantages and characteristics of offline expansion schemes. To realize the flexible participation of miner nodes and to avoid the risk of centralization, PRETRUST is not designed as an off-chain expansion scheme of transaction channel type but adopts the design of consortium chains with the structure of shard.

\section{Design of PRETRUST}\label{des}
\subsection{Model of blockchain}
Firstly, we introduce the basic concepts of the current cryptocurrency systems based on blockchains for more clear descriptions of PRETRUST. A blockchain system can be regarded as a kind of distributed database or distributed ledger maintained by multiple parties called “$miners$”. In a traditional blockchain network, a transaction is broadcasted to every miner and all the transactions which have not been recorded in the system constitute the “transaction pool”. In the data structure, a blockchain is a chain composed of blocks in which valid transactions are packed in a certain data structure like the Merkel Tree in the Bitcoin system. Blocks are linked in series by a hash pointer, which keeps the immutability of the whole system. Miners generate new blocks by verifying transactions in the transaction pool and packing them into new a new block, by which the chain will extend. Taking the Bitcoin system as an example, the speed of generating a new block is about costing 10 minutes. In the Bitcoin system, the time interval between generations of two adjacent blocks is limited by a lower bound \cite{garay2015bitcoin}.

\begin{figure*}[h]%
	\centering
	\includegraphics[width=0.9\textwidth]{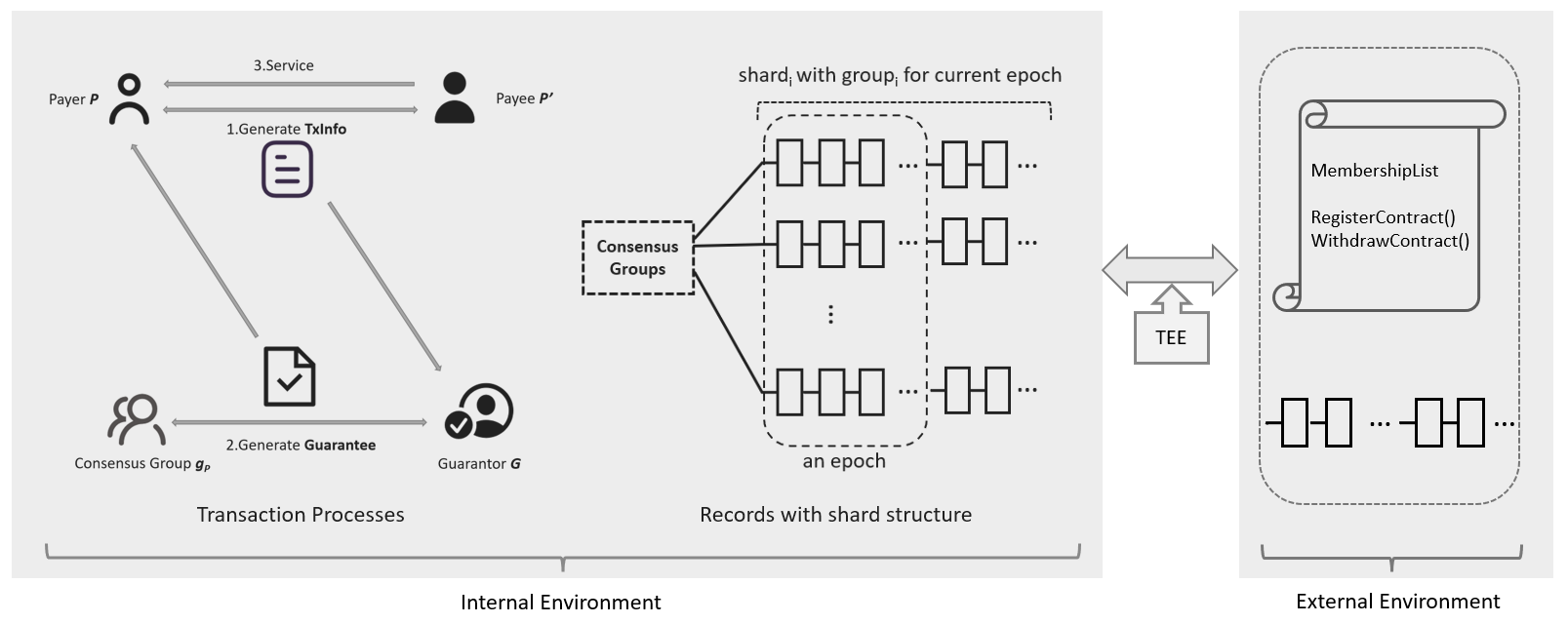}
	\caption{Key components of PRETRUST}\label{fig1}
\end{figure*}

Parties are identified by addresses (addr). An address originates from the public key of the key pairs ($pk$, $sk$) generated by parties themselves, and all the transactions (\textsf{tx}) are based on addr. In a transaction, for instance, payer $Alice$ with $addr_{A}$ intends to send c coins to payee $Bob$ with $addr_{B}$, $Alice$ is supposed to generate a transaction \textsf{tx} announcing that $c$ coins of $addr_{A}$ are transferred to $addr_{B}$. And the contents of \textsf{tx} need to be immutable with the digital signature from $Alice$ using her private key.

In a transaction, the key input is the coins in unspent transactions of $Alice$, Output is identified by $addr_{B}$ with the meaning of c coins from $Alice$ to $Bob$. Time is a delay parameter which means after Time blocks since the miner receives the transaction, \textsf{tx} will be packed into a block by miners. Data is a data field in which any data can be stored. And only when \textsf{tx} is signed by $Alice$, it is valid in blockchain systems.

\begin{table}[h]
	\begin{center}
			\caption{Transaction model}\label{tab1}%
			\begin{tabular}{llll}
				\toprule
				\multicolumn{4}{c}{Transaction \textsf{tx}} \\
				\midrule
				Input:  &&& $b$ coins of $Alice$   \\
				Output:  &&& $c$ coins to $Bob$   \\
				Time: &&& $time$   \\
				Data: &&& $arbitrary\ data$\\
				Signature: &&& \textsf{Sign(}$addr_{A}$, \textsf{tx)}\\
				\bottomrule
			\end{tabular}
	\end{center}
\end{table}

\subsection{Backbone of PRETRUST}
Protocols run in PRETRUST are relied on the secure parameter $\kappa$, which sets the details of rules mentioned in the following parts. Consider a transaction scenario in which three parties are involved: the payer $P$, the payee $P^{\prime}$, and the guarantor $G$. Here $G$ needs to be a member of the consensus group which keeps the ledger of $P$. In a transaction, $P$ needs to pay for $P^{\prime}$ to get the commodity. Under the PRETRUST framework, $P$ is supposed to find a trusted guarantor to generate a warrant declaring that $G$ will guarantee for $P$, and promise $P^{\prime}$ will get the payment for the transaction. Describe briefly for formal progress:

\begin{enumerate}
	\item The transaction information (\textsf{TxInfo}) is first generated by $P$ and $P^{\prime}$;
	\item $P$ sends \textsf{TxInfo} to the internal environment and \textsf{TxInfo} will be sent to the corresponding consensus group ($g_{P}$) which is responsible for the ledger of $P$;
	\item $G$ generates the current guarantee files (\textsf{PreGuarantee}) and gets the final warranty (\textsf{Guarantee}) with the verification from the corresponding consensus group ($g_{G}$). And \textsf{Guarantee} is supposed to contain the block expected information of the block which will record \textsf{Guarantee} in the future. And then $G$ sends it to $P$ after the verification from the consensus group ($g_{G}$) which is responsible for the ledger of $G$.
	\item $P^{\prime}$ receives \textsf{Guarantee} from $P$ and offers the commodity or service for $P$ after verification. This brief process is shown in Figure \ref{fig1} and the details are described in Section \ref{protocol}.
	
\end{enumerate}

As mentioned before, we use the “$external\ environment$” to denote public blockchain systems that PRETRUST relies on, and the “internal environment” to denote the elements which are off the public chain. Here are the main parts of the processing of PRETRUST:

\textbf{Interactions between the external environment and the internal environment:} To be compatible with the current currency system based on blockchain technology without reconstructing the entire blockchain architecture, PRETRUST is designed as a chain off the public blockchains with a shards structure. The internal environment is a framework of multiple shards working in parallel. Naturally, it is predominant to propose the interaction strategy between the external and internal environment. In PRETRUST, blockchain systems with the support of smart contracts are required. With the leverage of smart contracts, a list to record the membership of all the parties including guarantors and clients can be realized by distributed applications like notebooks based on contracts. Besides, with a smart contract design, any party can join in and leave PRETRUST. That is when a party needs to withdraw the deposit and cut the relationship with PRETRUST, he needs to use the corresponding smart contract.

Here, use \textsf{RegisterAccount()} to denote the smart contract for taking part in PRETRUST and \textsf{Withdraw()} to denote the smart contract for leaving PRETRUST. Since the internal environment and external environment are two isolated fields, to guarantee the security of invoking the contract \textsf{Withdraw()}, PRETRUST takes the leverage of TEE. TEE takes the information of accounts as its inputs and generates the \textsf{WithdrawlCertification} for the user who intends to withdraw the tokens. In summary, the interactions with the public chain system for PRETRUST are as follows: (1). Smart contract for being a member of PRETRUST by paying the deposit; (2). Smart contract for withdrawing the deposit and token in PRETRUST and aborting the membership. And these two smart contracts are both deployed on the public chain. The membership list generated by the smart contract will be the credential for the internal environment to update the information of taking part in and departure of any party. Besides, when a party intends to withdraw the token in the public chain, leverage of smart contracts is also needed to verify the validity of withdrawing permission like verifying the correctness of signatures based on the membership list. For all the parties in PRETRUST, the amount of the deposit paid to the contract account in the public chain will be converted into tokens in the internal environment. Correspondingly, any part willing to get tokens from the public chain needs to run an exchange execution first and get the output of the internal environment which is the input of the smart contract in the public chain. The specific implementation process and parameter format will be described in Section \ref{protocol}.

\textbf{Records of ledgers:} Ledgers in PRETRUST record different kinds of transactions that are distinctive from classical transaction structures. Ledgers are based on accounts like Ethereum, but to realize fast payment, protocols in PRETRUST are mainly designed for “guarantee-like” transactions. The main transaction information is called “\textsf{Guarantee}”, which will be packed into blocks by corresponding consensus groups. And the modification of the balance of a party is implemented via smart contracts with \textsf{Guarantee} as the inputs in the internal environment.

\textbf{Sharding strategy:} To handle transactions and supervision efficiently, PRETRUST is designed as a consortium chain  with a sharding structure in the internal environment. And there are two aspects of shard strategy required: shard of guarantors and transactions. The members of the consensus group of a certain shard are not static. And that means all guarantors in PRETRUST need to be reorganized periodically. The serial number of a shard is fixed while the members of the consensus group responsible for the ledger of this shard are dynamic. To realize dividing the guarantor randomly, reorganization relies on a random value. In PRETRUST, after a certain count of blocks: $epoch$, every shard will have published blocks $b = (b_{1}, b_{2}, …, b_{n})$ with the counter $n$. After the generation of $b_{n}$, the internal environment will begin the next epoch. All the guarantors received the end blocks $b_{n}$ of $epoch_{j}$, and they calculate a value of $epoch_{k}$: $globalHashk = \textsf{Hash(}b_{1}.hashValue \oplus b_{2}.hashValue  \oplus … \oplus b_{n}.hashValue\textsf{)}$. Division of guarantors in $epoch_{j+1}$ is based on $globalHash_{j}$ by computing: $shardNumber_{G} = \textsf{Upper(}n, globalHashk \oplus addr_{G}\textsf{)}$. Here, the notion “ $\oplus$ ” means xor computation, \textsf{Hash()} denotes hash value computation, \textsf{Upper(}$n$, $*$\textsf{)} denotes taking the upper $n$ bits of a bit sequence and aborting the rest. $globalHash_{k}$ can be regarded as a random number due to its generation method. And the randomization keeps each guarantor divided into a random shard and all the guarantors are reorganized in the same shard consist the new consensus group of this shard. With this kind of dynamic random reorganization, the risk of centralization can be avoided. On the other hand, the division of transactions is $shardNumber_{tx} = \textsf{Upper(}m, addr_{P}\textsf{)}$. Which shard a transaction will be arranged in is due to the address of the payer. The reason for the design is that it is efficient to keep the corresponding ledgers and make Guarantee for $P$. In one epoch, the corresponding consensus group is responsible for keeping the record of this shard and publishing blocks in the internal environment. And all guarantors need to keep the record of blocks from other shards since they may be divided into different consensus groups and shards.

\textbf{Election of guarantors:} There are two blockchain structures in the internal environment of PRETRUST: $Record Chain$ with shards structure and $Arbitration\ Chain$ with single chain structure. As mentioned before, when the consensus group $g_{P}$ receives a guarantee request with \textsf{TxInfo},  $g_{P}$ needs to calculate the values $guaranteePriority$ based on the addresses of all the group members and \textsf{TxInfo}.$id$. Here $guaranteePriority_{i} = add_{i} \oplus \textsf{TxInfo}.id$, then select the smallest $guaranteePriority_{i}$ and the candidate who guarantees for \textsf{TxInfo} is $G_{i}$ in $g_{P}$. When $G_{i}$ is absent or without any guarantee qualification, other members in $g_{P}$ can be the candidates according \textsf{guaranteePriorityList} in ascending order. And this mechanism can be realized by $g_{G}$: to sign the \textsf{PreGuarantee} received in a certain time according to the list.

\textbf{Supervision and punishment mechanism:} Each \textsf{Guarantee} is arranged to the corresponding shard according to the identity of its payer. The guarantor $G$ who generates certification \textsf{PreGuarantee} for $P$ is also the ledger keeper who is responsible for records of $P$. It is convenient for $G$ to check out the validation and illegality of the requirement from $P$ for a guarantee. After the guarantor $G$ makes the guarantee for a transaction, the amount of the remaining transactions the guarantor can guarantee will reduce by a certain account due to the value of the transaction. And this part of the token will not be unlocked unless $g_{G}$ receives the block containing the record of \textsf{Guarantee} from $g_{P}$. This is the criterion for $g_{G}$ to verify the validity of \textsf{PreGuarantee} (\textsf{PreGuarantee$_{1, 2}$}) and $g_{G}$ will sign \textsf{PreGuarantee} with group signature \cite{chaum1991group,camenisch1997efficient} if it is valid. Besides, the guarantee with multiple digital signatures is also a credential for $P^{\prime}$ to ensure his profits by calling the arbitration smart contract in the arbitration chain in the internal environment. Here arbitration chain is a blockchain consisting of one single chain structure with the proof of work consensus protocol, and it is maintained by all the guarantors. In the punishment mechanism of PRETRUST, once a guarantor and the corresponding consensus group $g_{G}$ are convicted, the cost of the punishment will be much more than the benefit gained from malicious behavior. $P^{\prime}$ can publish \textsf{Guarantee} signed by $G$ and $g_{G}$ to the arbitration chain by calling the arbitration smart contract to obtain the compensation. And before $P^{\prime}$ is tricked by $G$ by an illegal \textsf{Guarantee}, the corresponding consensus group $g_{G}$ which is responsible for the supervision of $G$ will check the validation of the \textsf{PreGuarantee}, and $g_{G}$ will not sign the \textsf{PreGuarantee}. Consequently, two checkpoints are ensuring the security of the payee in the supervision and punishment mechanism.

\textbf{Incentive mechanism:} In PRETRUST, there are two kinds of incentive mechanisms: reward for the guarantee on $Record\ Chain$ and reward for the record on $Arbitration\ Chain$. In \textsf{TxInfo}, $P$ needs to clarify the number of tokens paid as the guarantee fee. A part of the guarantee fee will be sent to $G$ and the rest will be sent to the members in $g_{G}$ who signed for the \textsf{PreGuarantee} after verification. The incentive mechanism of $Arbitration\ Chain$ is combined with the punishment mechanism. Once guarantors $G$ or the members of $g_{G}$ are convicted, a part of the tokens deducted will be sent to the guarantor who manages to generate the new block by proof of work consensus mechanism.

\section{Protocols of PRETRUST}\label{protocol}

We describe the details of PRETRUST in this chapter. There are mainly three phases in the process of implementing PRETRUST: participation, executing guarantee progress, and withdrawal.

\subsection{Participation}
As mentioned in Section III, any party who intends to take part in PRETRUST to deal with transactions or to be a guarantor needs to pay a deposit to the contract account $Account_{PRETRUST}$ on the public chain with the smart contract \textsf{RegisterAccount()}. For example, $Alice$, a miner of a public blockchain, intends to be a guarantor of PRETRUST framework. 

\begin{enumerate}
	\item $Alice$ first needs to pay the deposit to $Account_{PRETRUST}$ by invoking the smart contract \textsf{RegisterAccount()} with the deposit she intends to pay and user type as the inputs; 
	\item Public chain system run \textsf{RegisterAccount()}, after invoking the smart contract, the tuple \textsf{regInfo}$_{Alice}$ = ($addr_{Alice}$, $depositAmount$, $type$) will be added into a member list (\textsf{MembershipList}), which is a member of contract \textsf{RegisterAccount()}. Here $type$ = {“$guarantor$”, “$client$”}, $Alice$ can only be a normal client and deal with transactions with others if $type$ = “$client$” according to the inputs.
	
\end{enumerate}

\begin{table}[h]
	\begin{center}
		\begin{minipage}{174pt}
			\caption{Contents of \textsf{MembershipList}}\label{tab1}%
			\begin{tabular}{lll}
				\toprule
				\multicolumn{3}{c}{\textsf{MembershipList}} \\
				\midrule
				Address  & Type &  Deposit \\
				$addr_{PRETRUST}$  & $*$ & $*$  \\
				$addr_{Alice}$ & $guarantor$ & $a$ \\
				$addr_{Bob}$ & $client$ &  $b$\\
				$addr_{Charlie}$ & $client$ & $c$ \\
				\bottomrule
			\end{tabular}
		\end{minipage}
	\end{center}
\end{table}
\begin{algorithm}
	\renewcommand{\algorithmicrequire}{\textbf{Input:}}
	\renewcommand{\algorithmicensure}{\textbf{Output:}}
	\caption{\textsf{RegisterAccount()}}\label{algo1}
	\begin{algorithmic}[1]
		\Require $addr, deposit, type$
		\Ensure $register \ status$
		\If{$signatures$ are valid \textbf{and} $addr.balance$ $\geq$ deposit }\label{algln2}
		\State $addr$$\xrightarrow{deposit}$ PRETRUST 
		\State add $addr$ to \textsf{\textsf{MembershipList}}
		\State SUCCESS
		\Else
		\State FAILURE
		\EndIf
	\end{algorithmic}
\end{algorithm}

\subsection{Guarantee process}

In PRETRUST, the guarantee is the basic strategy. In this section, we give the generation and the entire process of the guarantee mechanism.
Generation and arrangement of \textsf{TxInfo}: In As the transaction scenario described in Section III. $P$ intends to deal with $P^{\prime}$ as the payer. First, they need to generate the \textsf{TxInfo}, which contains the contents of transaction information and signatures. After \textsf{TxInfo} is generated, it will be published by $P$ and finally sent to the corresponding shard with the current consensus group $g_{P}$. Group $g_{P}$ then selects the guarantor with the strategy described in Section \ref{des}. In \textsf{TxInfo}, \textsf{tx}$SN_{P}$ is the serial number of the counter of transactions from $P$.

\begin{center}
		\makeatletter\def\@captype{table}
		\caption{\textsf{TxInfo}}\label{tab1}%
		\begin{tabular}{ll}
			\toprule
			\multicolumn{2}{c}{\textsf{TxInfo} ($id$ = \textsf{Hash(}$m_{4},m_{5}$\textsf{)})} \\
			\midrule
			$m_{1}$  & $P$ pays $P^{\prime}$ c coins   \\
			$m_{2}$  & \textsf{tx}$SN_{P}$, guarantee fee   \\
			$m_{3}$ & \textsf{tx}$SN_{P^{\prime}}$   \\
			$m_{4}$ & \textsf{Sign(}$sk_{P}, (m_{1}, m_{2}$)\textsf{)}\\
			$m_{5}$ & \textsf{Sign(}$sk_{P_{\prime}}, (m_{1}\|m_{3}$)\textsf{)}\\
			\bottomrule
		\end{tabular}	
\end{center}

\begin{center}
		\begin{tabular}{ll}
			\toprule
			\multicolumn{2}{c}{\textsf{PreGuarantee$_{1}$}
			} \\
			\midrule
			$m_{1}$  & \textsf{TxInfo}   \\
			$m_{2}$  & $guarSN_{G}$, block information   \\
			$m_{4}$ & \textsf{Sign(}$sk_{P}, (m_{1}\|m_{2}$)\textsf{)}\\
			\bottomrule
		\end{tabular}
\end{center}

Processes of generating \textsf{Guarantee}: The generation of the final and valid guarantee needs three steps and two of them are temporary files.
\begin{enumerate}
	\item $P$ checks \textsf{TxInfo} and generates the first temporary guarantee file \textsf{PreGuarantee$_{1}$}
	, and then sends it to P;
	
	
	\item $P$ receives \textsf{PreGuarantee$_{1}$}
	and checks the correspondence of the shard. Then $P$ generates \textsf{PreGuarantee$_{2}$}
	with the signature for the contents of \textsf{PreGuarantee$_{1}$}
	and sends it to $G$;
		\item $G$ receives \textsf{PreGuarantee$_{2}$}
	and sends it to the corresponding shard of G, in which the current consensus group is $g_{G}$; 
	\item $g_{G}$ receives \textsf{PreGuarantee$_{2}$}
	and signs it with the group signature after verifying the validity of \textsf{PreGuarantee$_{2}$}
	, and finally, the valid guarantee file is generated. And then it will be sent to $P$ with the intermediary agent $G$. In \textsf{PreGuarantee$_{1}$}
	, $G$ is required to declare the expected block information in which the final guarantee file \textsf{Guarantee} will be contained.
	
\end{enumerate}	

\begin{center}
	\makeatletter\def\@captype{table}
	\caption{\textsf{PreGuarantee$_{2}$}
	}\label{tab1}%
		\begin{tabular}{ll}
			\toprule
			\multicolumn{2}{c}{\textsf{PreGuarantee$_{2}$}
			} \\
			\midrule
			$m_{1}$  & \textsf{PreGuarantee$_{1}$}
			\\
			$m_{2}$ & \textsf{Sign(}$sk_{P}, m_{1}$\textsf{)}\\
			\bottomrule
		\end{tabular}
\end{center}

\begin{center}
		\makeatletter\def\@captype{table}
	\caption{\textsf{Guarantee}}\label{tab1}%
	\begin{tabular}{ll}
		\toprule
		\multicolumn{2}{c}{\textsf{Guarantee}} \\
		\midrule
		$m_{1}$  & \textsf{PreGuarantee$_{2}$}
		\\
		$m_{2}$ & \textsf{Sign(}$sk_{g}, m_{1}$\textsf{)}\\
		\bottomrule
	\end{tabular}
\end{center}

%
	
	\begin{algorithm}
		\renewcommand{\algorithmicrequire}{\textbf{Input:}}
		\renewcommand{\algorithmicensure}{\textbf{Output:}}
		\caption{\textsf{VerifyPreGuarantee$_{1}$()}
		}\label{algo1}
		\begin{algorithmic}[1]
			\Require $\textsf{PreGuarantee$_{2}$}
			$
			\Ensure $\textsf{Guarantee}$
			\If{$signatures$ are valid \textbf{and} no other prior guarantee response}\label{algln2}
			\State sign \textsf{PreGuarantee$_{1}$}
			
			\State generate \textsf{PreGuarantee$_{2}$}
			with the signature of $P$
			\State SUCCESS
			\Else
			\State FAILURE
			\EndIf
		\end{algorithmic}
	\end{algorithm}

	\begin{algorithm}
		\renewcommand{\algorithmicrequire}{\textbf{Input:}}
		\renewcommand{\algorithmicensure}{\textbf{Output:}}
		\caption{\textsf{GenerateGuarantee()}}\label{algo1}
		\begin{algorithmic}[1]
			\Require $\textsf{PreGuarantee$_{2}$}
			$
			\Ensure $\textsf{Guarantee}$
			\If{$signatures$ are valid \textbf{and}\\
				$guaranteePriority$ is valid \textbf{and} \\
				available deposit of $G$ $\ge ( c + guarantee fee ) * \kappa $ \textbf{and} \\
				$guarSN$  is valid}\label{algln2}
			\State sign \textsf{PreGuarantee$_{2}$}
			
			\State generate \textsf{Guarantee} with group signature of $G^{\prime}$
			\State SUCCESS
			\Else
			\State FAILURE
			\EndIf
		\end{algorithmic}
	\end{algorithm}
	

	\begin{algorithm}
	\renewcommand{\algorithmicrequire}{\textbf{Input:}}
	\renewcommand{\algorithmicensure}{\textbf{Output:}}
	\caption{\textsf{VerifyTxInfo()}}\label{algo1}
	\begin{algorithmic}[1]
		\Require $\textsf{TxInfo}$
		\Ensure \textsf{PreGuarantee$_{1}$}
		\If{$signatures$ are valid \textbf{and} $P.balance$ $\geq$ $c + guarantee.fee$ deposit \textbf{and} \textsf{tx}$SN$ is valid }\label{algln2}
		\State sign \textsf{TxInfo}
		\State generate \textsf{PreGuarantee$_{1}$} with the signature of $G$
		\State SUCCESS
		\Else
		\State FAILURE
		\EndIf
	\end{algorithmic}
\end{algorithm}

Transactions between $P$ and $P^{\prime}$: $P$ finally receives Guarantee and shares it with $P^{\prime}$. $P^{\prime}$ will offer a commodity or service to $P$ after verifying the validity of \textsf{Guarantee}.

\subsection{Withdrawal process}
In PRETRUST, when a user intends to withdraw tokens deposited on public blockchains or the balance in the internal environment, he needs to take the leverage of TEE. In the model of TEE of this paper, TEE needs to generate its key paris ($pk_{PRETRUST}$, $sk_{PRETRUST}$) first and $pk_{PRETRUST}$ needs to be an initial parameter of \textsf{MembershipList}. Information on the balance needs to be available as inputs for TEE so that a valid \textsf{WithdrawalCertification()} can be generated. For example, $Alice$ intends to withdraw c coins from her balance.

\begin{enumerate}
	\item She first needs to publish a withdrawal request \textsf{WithdrawalRequest} announcing the amounts of tokens;
	
	\item \textsf{WithdrawalRequest} will be sent to the corresponding consensus group $g_{A}$. Then $g_{A}$ verifies the validation of \textsf{WithdrawalRequest} and generates \textsf{WithdrawalCheck} if there is nothing illegal. And $g_{A}$ needs to deduct the corresponding tokens of $Alice$ in the internal environment instantly; 
	\item $g_{A}$ sends \textsf{WithdrawalCheck} to TEE and TEE publishes it instantly in the internal environment. Then, the statement of $Alice$ will be locked and all the nodes in PRETRUST will refuse to deal with transactions with $Alice$ temporarily; 
	
	\item TEE checks the information of \textsf{WithdrawalCheck} with the records of the internal environment and returns to \textsf{WithdrawalCertification} with the time stamp and signature of TEE. After this step, its “$lock\ state$” of $Alice$ will be relieved; 
	\item $Alice$ invokes contract \textsf{Withdraw()} with \textsf{WithdrawalCertification} as inputs and gets the deposit.

\end{enumerate}

	\begin{table}[h]
	\begin{center}
			\caption{\textsf{WithdrawalRequest}}\label{tab1}%
			\begin{tabular}{ll}
				\toprule
				\multicolumn{2}{c}{\textsf{WithdrawalRequest}} \\
				\midrule
				$m_{1}$  & $addr$ $\coloneqq$ address of $Alice$   \\
				$m_{2}$ & $token$ $\coloneqq$ amounts $Alice$ intends to withdraw \\
				$m_{3}$  & \textsf{Sign(}$sk_{Alice}$, ($m_{1}\|m_{2}$))   \\
				\bottomrule
			\end{tabular}
	\end{center}
\end{table}

	\begin{table}[h]
	\begin{center}
		\makeatletter\def\@captype{table}
		\caption{\textsf{WithdrawalCertification}}\label{tab1}%
		\begin{tabular}{ll}
			\toprule
			\multicolumn{2}{c}{\textsf{WithdrawalCertification}} \\
			\midrule
			$m_{1}$  & $time$ $\coloneqq$ in an interval of time interval  \\
			$m_{2}$ & $addr$ $\coloneqq$ address of $Alice$  \\
			$m_{3}$ & $token$ $\coloneqq$ tokens to withdraw \\
			$m_{4}$  & \textsf{Sign(}$sk_{Alice}$, ($m_{1}\|m_{2}$)\textsf{)}   \\
			\bottomrule
		\end{tabular}
	\end{center}
\end{table}


\section{Analysis of security}

In general, PRETRUST offers another access for miners in blockchains to gain rewards for keeping records of blocks. And as many miners join in and maintain the internal environment, the robustness of the system can be enhanced. Moreover, PRETRUST limits the maximum of transactions that a guarantor can guarantee to ensure that a guarantor cannot get an unlimited guarantee reward, in which case the risk of centralization can be avoided again.  In this section we mainly discuss the possible malicious behaviors in PRETRUST and the protection mechanisms facing these situations:  

\begin{itemize}
	\item \textbf{No privilege for a guarantor:} In PRETRUST, if a party of a consensus group $G$ intends to guarantee a transaction, he will play a role as a normal guarantor. This means all the behaviors are supervised by the corresponding consensus group via the strict generation of the guarantee file and supervision mechanism from the internal environment. In this case, there is no privilege or trapdoor for him.
	
	\item \textbf{No benefits for refusing to offer transaction results:} For a payee $P^{\prime}$, if he wants to obtain benefits after delivering the commodity, the only way is to broadcast the final generated \textsf{Guarantee} with his signature. In this case, the consensus group g$P^{\prime}$ will update the record $P^{\prime}$ and the added balance of $P^{\prime}$ is valid. 
	
	\item \textbf{\textsf{TxInfo} for guarantee requirement:} After $P$ and $P^{\prime}$ generate \textsf{TxInfo}, the corresponding shard with the temporary consensus group $g_{P}$ will be responsible for the verification of \textsf{TxInfo}. Notice $g_{P}$ is the keeper of the ledger of $P$, and it is more efficient to check out whether $P$ can afford this transaction due to the current balance of $P$. If $P$ is malicious, $g_{P}$ will be aware of that and refuse to make a guarantee for $P$.
	\item \textbf{\textsf{PreGuarantee}:} The final valid \textsf{Guarantee} needs the verification of the corresponding shard with temporary consensus group $g_{G}$. $g_{G}$ is responsible for the ledger of $G$ in the current epoch. And $g_{G}$ will check out the malicious behaviors of $G$ (such as not enough balance of the deposit) and refuse to generate the final guarantee file for $G$ on this illegal occasion. In another situation: Once $P^{\prime}$ receives the final guarantee file \textsf{Guarantee}, $P^{\prime}$ is supposed to offer the commodity to $P$ and end up the transaction. If $P^{\prime}$ finds out at last that there is no record of \textsf{Guarantee} according to the information of \textsf{Guarantee} he received, $P^{\prime}$ invokes the smart contract \textsf{Arbitration()} on the $Arbitration\ Chain$ to get the compensation. And after invoking successfully, $G$ will be punished.
	
	\begin{algorithm*}
		\renewcommand{\algorithmicrequire}{\textbf{Input:}}
		\renewcommand{\algorithmicensure}{\textbf{Output:}}
		\caption{\textsf{Arbitration()}}\label{algo1}
		\begin{algorithmic}[1]
			\Require $\textsf{PreGuarantee}$
			\Ensure $ArbitrationStatus$
			\If{$signatures$ are valid \textbf{and}
				$blockContents$[\textsf{Guarantee}.block information] $\neq$ $GuaranteeContents$
			}\label{algln2}
			\State \textsf{Guarantee}.$G$ $\xrightarrow{\textsf{tx}Value* \kappa.CompensationWeight\ {\rm tokens}}$ \textsf{Guarantee}.$P^{\prime}$
			\State \textsf{Guarantee}.$G$ $\xrightarrow{\textsf{tx}Value* \kappa.PunishmentWeight\ {\rm tokens}}$ block generator 
			\State SUCCESS
			\Else
			\State FAILURE
			\EndIf
		\end{algorithmic}
	\end{algorithm*}

	\item \textbf{\textsf{WithdrawalRequest}:} $P$ needs to get \textsf{WithdrawalRequest} from TEE via $G$, and the security of TEE guarantees the security of the action of withdrawal.
	
	\begin{algorithm*}
		\renewcommand{\algorithmicrequire}{\textbf{Input:}}
		\renewcommand{\algorithmicensure}{\textbf{Output:}}
		\caption{\textsf{Withdraw()}}\label{algo1}
		\begin{algorithmic}[1]
			\Require \textsf{WithdrawalCertification}, $\kappa$
			\Ensure $withdraw\ status$
			\If {\textsf{WithdrawalCertification}.$signature$ is valid \textbf{and}
				$System.time – \textsf{WithdrawalCertification}.time \leq \kappa.timeInterval$}\label{algln2}
			\State PRETRUST $\xrightarrow{\textsf{WithdrawalCertification}.token}$ \textsf{WithdrawalCertification}.$addr$
			\State SUCCESS
			\Else
			\State FAILURE
			\EndIf
		\end{algorithmic}
	\end{algorithm*}
\end{itemize}

\section{Discussion}
There is no doubt that the original POW consensus algorithm of the blockchain proves that consensus algorithm and single chain structure are powerful strategies to ensure security among parties that do not trust each other. However, the cost of doing so is a great sacrifice in efficiency. Graph-based blockchain systems, such as DAG \cite{churyumov2016byteball,silvano2020iota}, proposes the idea of parallelism to solve the efficiency problem rather than the single chain structure, but the complex design of graphs will bring new problems in security and consistency. The subsequent Lighting Network \cite{poon2016bitcoin} and Raiden Network, taking these drawbacks into full consideration, design an on-off combination mode to transfer more interactions between users off the public chain, and solved the problems of duplication and consistency in transaction logic through clever design. However, this kind of design relies too much on the central node, and the disadvantage of node reorganization makes the risk of centralization increase with the expansion of the network scale. Besides, the channel plays an important role in this kind of design for improving efficiency, but channels also bring limitations in operating in different trade groups. 

In PRETRUST, the idea of on-off is followed. More importantly, the protocol uses a pre-trust mechanism to further optimize the transaction process. Besides, to avoid potential centralization risks, PRETRUST adopts a dynamic grouping design. And the adoption of the strategy of dividing consensus groups and users will not produce transaction islands, causing the convenience of transactions to be only reflected in relatively fixed patterns. Periodic reorganization and flexible admission mechanisms improve the security and robustness of the system. The comparison is shown in the Table \ref{compare}.
\begin{table}
	\begin{center}
			\caption{Comparison with different schemes}\label{compare}%
			\begin{tabular}{lcccc}
				\toprule
				Scheme & Structure & Non-Centralization & Optimization & Flexibility\\
				\midrule
				BitCoin\cite{nakamoto2008peer} & on & \ding{51} \ding{51}& \ding{53} & \ding{53}\\
				DAG\cite{churyumov2016byteball,silvano2020iota} & on & \ding{51} & \ding{51} & \ding{51}\\
				Lightning\cite{poon2016bitcoin} & on-off & \ding{53} & \ding{51} \ding{51} & \ding{51} \ding{51}\\
				Raiden\cite{hees2016raiden} & on-off & \ding{53} & \ding{51} \ding{51} & \ding{51} \ding{51}\\
				PRETRUST & on-off & \ding{51} & \ding{51} \ding{51} & \ding{51} \ding{51}\\
				\bottomrule
			\end{tabular}
			
	\end{center}
\end{table}

\section{Conclusions}
To improve the efficiency of payment systems based on blockchains, this paper proposes PRETRUST, a scheme to realize fast payments. The main motivation of the framework is to introduce a third guarantor to the certificate for a transaction in advance to save the waiting time for the payer and the payee. PRETRUST is running on top of current blockchains which support smart contracts. Though PRETRUST is a kind of off-chain strategy to solve the problem of efficiency of current public blockchain systems, the internal designs are based on blockchain structure. With this method, PRETRUST turns out to be a method to enhance efficiency with security and compatibility. For a miner node in public blockchains, the threshold for him to join the internal environment is low, which makes it possible for miners to join the system with freedom. And this makes the system more robust. Finally, we formally analyze the security of the general framework by considering the kinds of malicious behaviors of parties. In the whole transaction process, multiple-verification is involved to ensure the security of the transaction.

\bibliographystyle{ieeetr}
\bibliography{references}
\end{document}